\documentclass[10pt,letterpaper,showkeys,superscriptaddress,sort&compress,rmp,round,floatfix]{revtex4-1}
\usepackage[dvips]{geometry}
\usepackage{snapshot}
\usepackage{mymanuscript}
\begin{document}

\title{Evolution of male life histories and age-dependent sexual
  signals under female choice \footnote{Copyright 2013 Joel
    J. Adamson.  Licensed under Creative Commons
    Attribution-Noncommercial-ShareAlike License.}  \footnote{To be
    published in PeerJ; http://dx.doi.org/10.7717/peerj.225}
  \footnote{Code and data available as Adamson, Joel (2013): Evolution
    of Age-Dependent Traits. figshare.
    http://dx.doi.org/10.6084/m9.figshare.783068}}

\author{Joel J. Adamson \\
  Ecology, Evolution and Organismic Biology\\CB \#3280 \\University of
  North Carolina, Chapel Hill, NC
  27599\\919-843-2320\\adamsonj@ninthfloor.org}

\bibliographystyle{chicago}
\begin{abstract}
  Sexual selection theory models evolution of sexual signals and
  preferences using simple life histories.  However, life-history
  models predict that males benefit from increasing sexual investment
  approaching old age, producing age-dependent sexual
  traits. Age-dependent traits require time and energy to grow, and
  will not fully mature before individuals enter mating competition.
  Early evolutionary stages pose several problems for these
  traits. Age-dependent traits suffer from strong viability selection
  and gain little benefit from mate choice when rare.  Few males will
  grow large traits, and they will rarely encounter choosy females.
  The evolutionary origins of age-dependent traits therefore remain
  unclear. I used numerical simulations to analyze evolution of
  preferences, condition (viability) and traits in an age-structured
  population. Traits in the model depended on age and condition
  (``good genes'') in a population with no genetic drift.  I
  asked \begin{inparaenum}[(1)]
  \item if age-dependent indicator traits and their preferences can
    originate depending on the strength of selection and the size of
    the trait;
  \item which mode of development (age-dependent versus
    age-independent) eventually predominates when both modes occur in
    the population;
  \item and if age-independent traits can invade a population with
    age-dependent traits.
  \end{inparaenum}
  Age-dependent traits evolve under weaker selection and at smaller
  sizes than age-independent traits.  This result held in isolation
  and when the types co-occur.  Evolution of age-independent traits
  depends only on trait size, whereas evolution of age-dependent
  traits depends on both strength of selection and growth
  rate. Invasion of age-independence into populations with established
  traits followed a similar pattern with age-dependence predominating
  at small trait sizes.  I suggest that reduced adult mortality
  facilitates sexual selection by favoring the evolution of
  age-dependent sexual signals under weak selection.
\end{abstract}
\maketitle

\section{Introduction}

\label{sec:intro}

Sexual selection theory studies the exaggeration of secondary sexual
traits and corresponding preferences in the opposite sex.  The most
well-developed portion of this theory describes maintenance of female
preferences under indirect (genetic) benefits \citep{jones2009mate}.
The theory of indicator traits explains the maintenance of female
preferences with genetic correlations between male viability and
female preferences \citep{kokko06:_unify}.  Testing good genes theory
requires attention to life-history variables \citep{kokko2001faH}.
Experimenters often measure correlations between health of sires and
survival of offspring
\citep{evans_divergent_2011,jacob_male_2007,jacob_effects_2010,kokko2002ssc}.
Secondary sexual traits, particularly in large vertebrates, require
time to grow and young males can enter mating competition against
older males with larger weapons \citep{pemberton04:_matin}.  Older
males provide superior genetic benefits in some cases
\citep{Brooks2001308}.  Age-dependent traits occur frequently in
nature and form frequent subjects for laboratory studies of sexual
selection and coercion \citep{bonduriansky_reproductive_2005}.
However, most sexual selection models do not account for the growth of
traits and include only simplified life-histories
\citep{kokko06:_unify,kokko2001faH,kokko2002ssc}.

Life-history theory suggests that sexual selection theory could
benefit from modeling more complex life-histories.  Low adult
mortality leads to a stable strategy of age-dependent male
reproductive effort.  \citet{kokko97:_evolut_stabl_strat_of_age} found
evolutionarily stable strategies for age-dependent strategies under
fairly broad conditions.  \citet{proulx2002oms} found that males
benefit from increasing mating effort as they age and reproductive
opportunities decline.  They predicted that condition should be
positively correlated with delays in investment.  High condition males
signal more at older ages.  A third, more recent study employing
similar techniques found that optimal higher-quality males will
postpone trait growth until the onset of breeding
\citep{rands_dynamics_2011}.

Life-history models and long-term studies of vertebrates suggest a
specific class of secondary sexual traits that require explanation.  I
define age-dependent traits as ``quantitative traits'' that males grow
throughout their reproductive lifetimes.  Antlers of deer
\citep{kodric-brown_truth_1984}, horns on sheep
\citep{pemberton04:_matin,Coltman2002} and body size in pinnipeds
\citep{clinton93:_sexual_selec_effec_male_life} and primates
\citep{courtiol_natural_2012,Geary2002,Mace20001} form good examples
of age-dependent morphological traits.  Others examples come from
behavioral traits that change over the lifetime due to experience or
aging, such as song repertoires \citep{Hiebert1989266,Gil2001689},
nest building \citep{EVANS1997749}, performance ability
\citep{Judge2011185,verburgt_male_2011,Ballentine2009973,garamszegi_agedependent_2007},
and social connectivity
\citep{oh_structure_2010,mcdonald_cooperative_1994}.  I define
age-independent traits as morphological patterns that are relatively
stable over the lifetime, and qualitatively different from juvenile
morphology at the time of breeding.  Readers could refer to
age-independent traits as ``qualitative traits.''  Plumage patterns in
birds (e.g.\ ``breeding plumage'') provide examples of age-independent
traits.  Some plumage patterns do change over the reproductive
lifespan \citep{evans_divergent_2011} making them age-dependent by the
definition above.

The evolutionary origins of these age-dependent sexual signals and
their corresponding preferences remain unclear, despite the clarity of
strategic models.  First, when traits require a large investment of
resources to grow, they require time to mature.  Such a trait would
require reduced adult mortality.  A strongly age-dependent trait could
suffer selection before it grows enough to make it attractive.
Second, frequency dependence crucially affects the origin of costly
sexual signals \citep{KP:82}.  If we suppose that both trait and
preference arise by mutation, then males growing age-dependent traits
will rarely encounter choosy females during early evolutionary stages.
Sexual selection will have limited opportunity to increase the trait.
Such a trait could be eliminated by selection, or lost to drift in
finite populations.  This introduces a critical time period the trait
must survive before proceeding to fixation.  Third, signal honesty
poses another problem.  For some kinds of traits, males will have
similar trait values at young ages regardless of variation in
condition.  Age-dependence of traits thereby weakens both the
heritability of the trait and the phenotypic correlation between the
trait and condition.  Finally, at the evolutionary origin of a trait,
only a subset of strategies occurs in the population.  Stability of a
particular strategy does not tell us which will predominate under such
a restricted strategy set.  This is a classic problem in game theory,
with a voluminous literature \citep[see][and references
therein]{Hammertime}.

A model of evolutionary dynamics would balance the theory of
age-dependent traits.  I used numerical simulations to investigate the
evolution of an age-dependent trait under sexual selection.  All males
start with similar trait values and grow their traits in a
condition-dependent manner.  The model population consists of haploid,
dioecious (male and female) individuals.  Selection on the male trait
produces the age structure of the population.  For the purposes of the
model, age-dependent and age-independent traits differ in that
age-independent traits are fully grown at the time of first breeding,
and only vary between males due to condition-dependence.  I ask three
specific questions.  First, how does the strength of selection and the
growth rate of the trait affect its evolution?  Second, I investigate
which mode of trait development (age-dependent versus age-independent)
will eventually predominate when both are initially present and rare.
Third I ask if age-independence can invade a population with
established age-dependent traits and preferences.  I investigate these
questions by examining the differences between conditions for
evolution of age-dependent and age-independent traits.  I show that
strongly age-dependent traits can increase in frequency at smaller
sizes than age-independent traits, both in the presence and absence of
age-independent traits.  Age-dependent traits in my model require weak
selection to eventually predominate.  My results suggest that weak
selection, strong age-dependence and reduced adult mortality can lead
to the exaggeration of sexual signals in species with extended
lifespans.  This suggests age-dependent traits are compatible with
life-histories seen in long-lived vertebrates.

\section{Model}
\label{sec:model}
\begin{figure*}
  \centering
  \includegraphics[width=150mm]{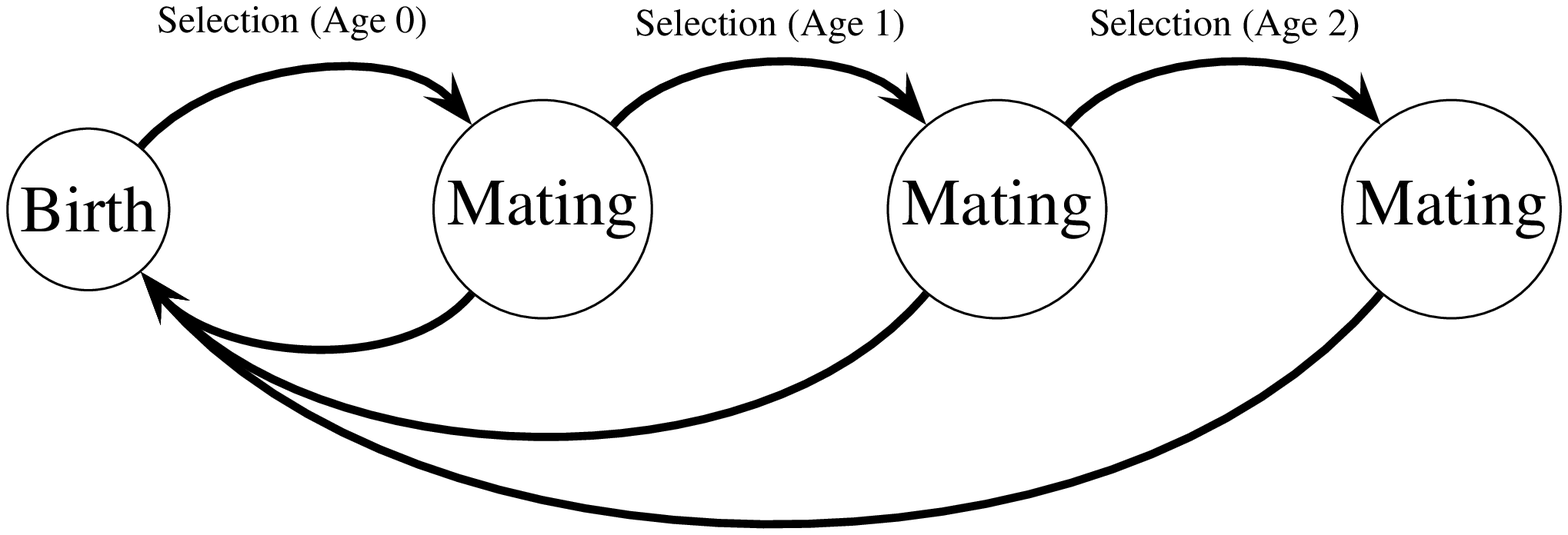}
  \caption[Life cycle]{Haploid life cycle for cohorts of males: haploid males
    emerge, grow, undergo selection, then mating, then repeated
    episodes of selection and mating.  Meiosis then produces new
    zygotes following mating.  Mutation in condition loci occurs
    before selection.}
  \label{fig:lcycle}
\end{figure*}
My model extends a classic model of female choice \citep{KP:82} by
adding age-dependence, iteroparity, overlapping generations of males,
and condition-dependence.  The model uses a haploid life cycle
(Figure~\ref{fig:lcycle}): new zygotes arise from meiosis and undergo
viability selection followed by mating.  Males that survive the first
round of selection and mating proceed to another round of selection
and mating, and so on.  Males provide no direct benefits to females,
and can mate with multiple females.  Each female mates once and lives
for one episode of viability selection followed by mating.  I assume a
large population size so that we can ignore the effects of genetic
drift.  All male mortality results from selection on the trait and
condition phenotypes except for males in the terminal age class, who
are removed from the simulation unconditionally (see
Equation~(\ref{eq:moorad})).  Similarly, females do not suffer costs
of choice, and all selection on females reflects selection on
condition.

The model genome consists of five diallelic loci: two ``condition'' or
``intrinsic viability'' loci ($\clocus$), the trait locus ($\tlocus$),
the preference locus ($\plocus$) and the age-dependent mode of
expression locus ($\flocus$).  I refer to alleles of interest, e.g.\
the preference or trait allele, or beneficial condition alleles, with
the subscript $2$, and denote their frequencies by lower-case $p$ with
a subscript corresponding to the locus.  For example $\pp$ represents
the frequency of the choosiness allele $\ptwo$.  The alleles at the
condition loci are either ``beneficial'' or ``deleterious.''  The
number of beneficial alleles across loci adds up to an individual's
condition phenotype $C$.  Biased mutation from beneficial to
deleterious ($C_{2}\rightarrow C_{1}$) occurs in condition loci at a
rate of $0.001$ per individual per generation; mutation occurs in the
zygote stage, before the first round of selection (see
Figure~\ref{fig:lcycle}).

\begin{figure*}
  \includegraphics[width=150mm]{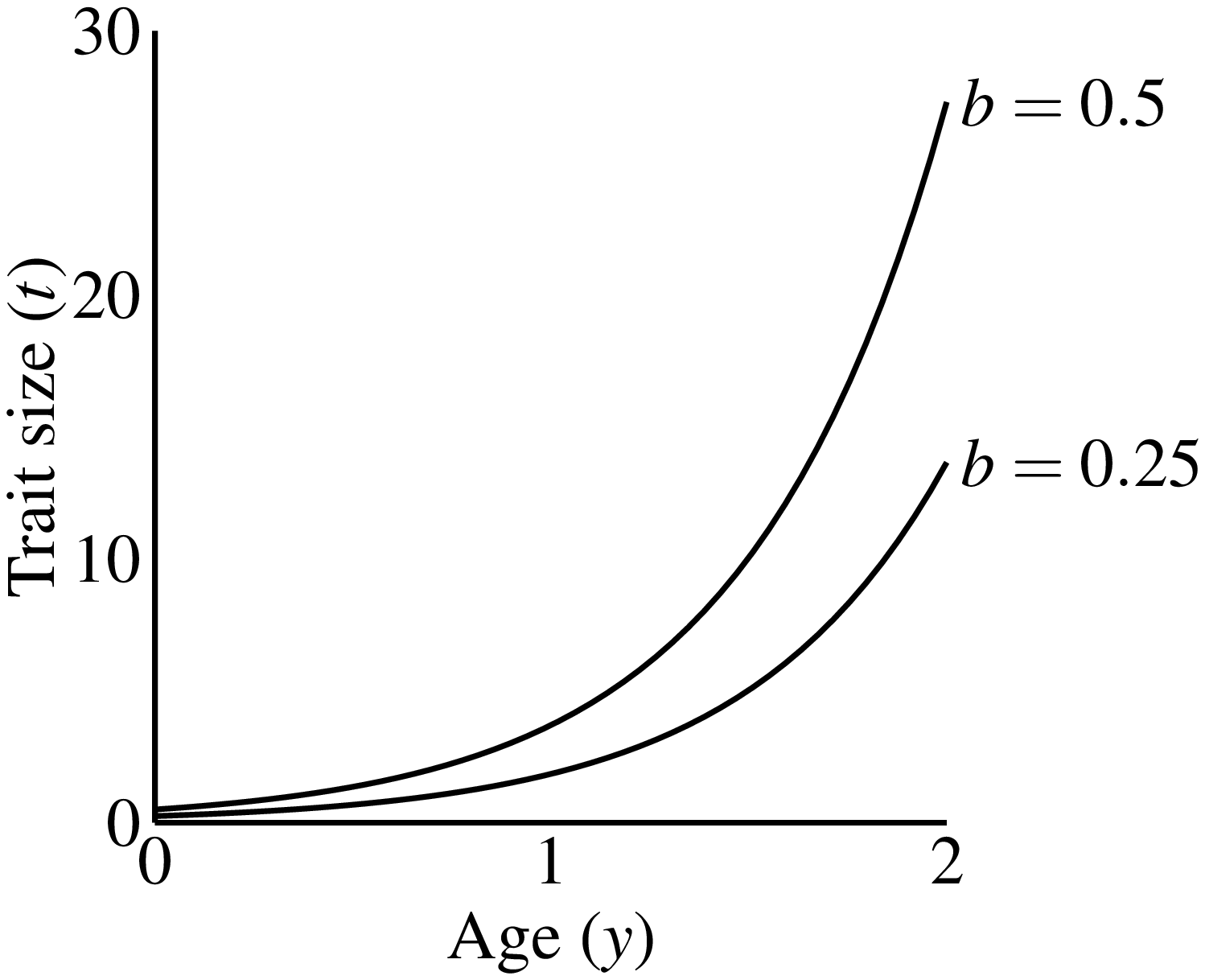}  
  \caption[Male trait growth at varying condition]{Male trait growth
    for high-condition males at two values of the growth coefficient,
    $b = 1.0$ and $b = 0.5$.  Trait size at age $0$ equals $b$.}
  \label{fig:trait}
\end{figure*}

Males carrying the $\tone$ allele do not produce the trait, regardless
of condition.  Males produce the trait if they have the trait allele
$\ttwo$.  A male aged $y$ with condition phenotype $C$ carrying
$\ttwo$ has trait size
\begin{equation}
  \label{eq:btrait}
  t (C, y) =  b e^{Cy}.
\end{equation}
where $b$ is a parameter controlling the size of the trait, that I
refer to as the ``growth coefficient.''  I chose this exponential
function to emphasize three characteristics:
\begin{inparaenum}[(1)]
\item all males display the same trait size at age $0$, as long as
  they carry the trait allele;
\item a large disparity in size between young and old males; and
\item simple scaling of the male trait size via the growth coefficient ($b$).
\end{inparaenum} Other trait functions do occur in nature
\citep[see][]{johnson_sexual_2011,poissant08:_quant}, and could have
different consequences for evolutionary dynamics (see
\titleref{sec:discussion}).  Figure~\ref{fig:trait} shows that
age-specific trait size linearly depends on the growth coefficient
(the constant $b$).  The growth coefficient linearly corresponds to
the size of the trait across constant age and condition.  A larger
value of $b$ in a particular population (i.e.\ simulation) signifies
that males of a particular genotype attain larger trait values than
they would in populations with smaller $b$-values.

The $\plocus$ locus controls mate choice behavior of females.  Mate
choice occurs by relative preference \citep[as in][]{KP:82} as a
function of trait size.  Females carrying $\ptwo$ encountering a male
with trait size $t$ are $ \phi (t) = 1 + \alpha{} t$ times more likely
to mate than non-choosy females.  Non-choosy females, carrying
$\pone$, mate randomly ($\phi (t) = 1$).  The mating process
normalizes female mating frequency such that all females have equal
mating success (see Equation~\ref{eq:mtable}).  In other words, all
females mate in each mating cycle.  Choosy females do not suffer any
viability or opportunity costs.

The $\flocus$ locus controls mode of development.  Males carrying the
$\fone$ allele show age-dependent expression, whereas carriers of the
$\ftwo$ allele express the trait throughout their lives at at one of
three levels in a particular simulation:
\begin{inparaenum}[(1)]
\item $t(C,0) = b$, the trait value of a $0$-year-old;
\item $t(C, y_{max})$, the trait value of the oldest males in the
  population (still dependent on condition); or
\item \label{item:tbar} $\bar{t}$, the population mean trait value.
\end{inparaenum}
The third set of simulations sought to create a population where
age-independent ($\ftwo$) males were of intermediate attractiveness,
between young or unornamented males and older, age-dependent males.
Fixation of age-dependence in this case shows that significant mating
advantages accrue later in life, despite a period of lesser
attractiveness early in life.  Fixation of age-independence, on the
other hand, would show that early-life attractiveness and costs were
more important than potential gains from mating later in life.
Contrast this with the second set of simulations: age-dependent males
only reach the attractiveness of their age-independent counterparts,
when they reach the final age class ($y_{max}$).  Age classes run from
$0$ (youngest) to $y_{max}$.  A youngest age class of $0$ conveniently
yields young males the trait size of $b$ in
Equation~(\ref{eq:btrait}), as well as using the same indexing
convention as the computer simulation.  The number of age classes in
the population is $y_{max}+1$.

For the third set of simulations, where $\ftwo$ males carried $t =
\bar{t}$, I updated $\bar{t}$ in every iteration.  My goal was
ensuring that a class of males of intermediate attractiveness
persisted in the population.  Therefore all males (regardless of
condition) carrying $\ftwo$ received $\bar{t}$ as their trait value
for a particular episode of mating, then I updated their values in the
next iteration, following changes in $\bar{t}$.  
Furthermore, the value of $\bar{t}$ used in these simulations
reflected the full range of variation in condition, despite the lack
of condition-dependence for $\ftwo$ males.  I therefore calculated the
mean trait as
\begin{equation}
  \label{eq:tbar}
  \bar{t} = \frac{\sum_{t=0}^{t_{max}}t f(t,y)}{y_{max}+1}
\end{equation}
where $f(t,y)$ describes the frequency of males with trait value $t$
at age $y$ over $y_{max}+1$ age classes.  The average is taken over
all males, from unornamented males to the maximum trait size of
$t_{max} = b e^{\mathcal{C}y_{max}}$, where $\mathcal{C}$ represents
the largest possible number of condition alleles (i.e. number of
condition loci).  Males carrying $\ftwo$ contributed to the population
mean as if their traits were age-dependent, i.e.\ contributing $t
(C,y) = b e^{Cy}$ to the calculation in Equation~(\ref{eq:tbar}).
If the trait allele ($\ttwo$) were to spread, then a contribution of
$\bar{t}$ by $\ftwo$ males would depress the trait value for $\ftwo$
males and produce a delay in following phenotypic changes.  Although
these procedures reduced biological realism from an individual
perspective, they maintained the population genetic conditions
relevant to the question at hand.

I used one level of expression per simulation, including
simulations where the male population was fixed for $\ftwo$ (an
age-independent population).  In a different set of simulations
$\ftwo$ initially occurred at a low frequency, comparable to the
frequency of the trait, so that age-independent and age-dependent
expression were in competition.

\begin{figure*}
  \centering
  \subfloat[$\mu=10$,$\nu=1$]{\label{subfig:wsurfone}
    \includegraphics[width=58mm]{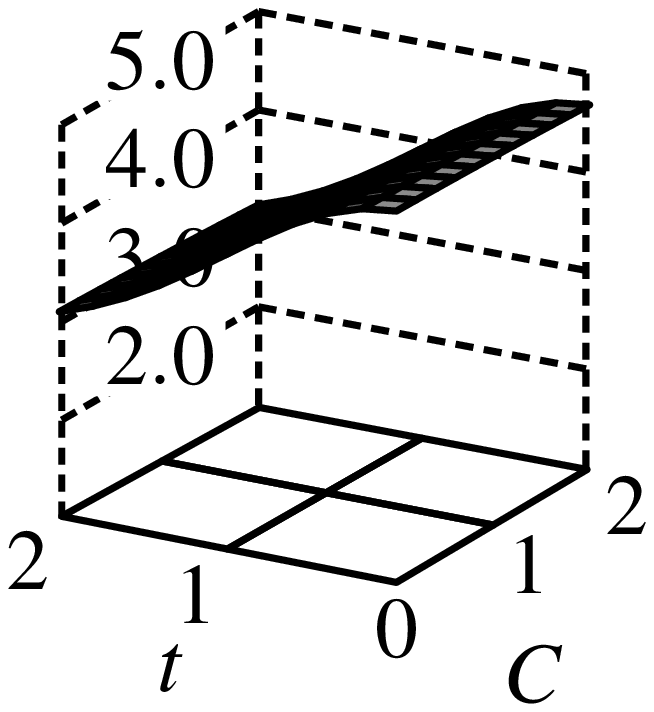}}
  \subfloat[$\mu=10$,$\nu=0.1$]{\label{subfig:wsurftwo}
    \includegraphics[width=58mm]{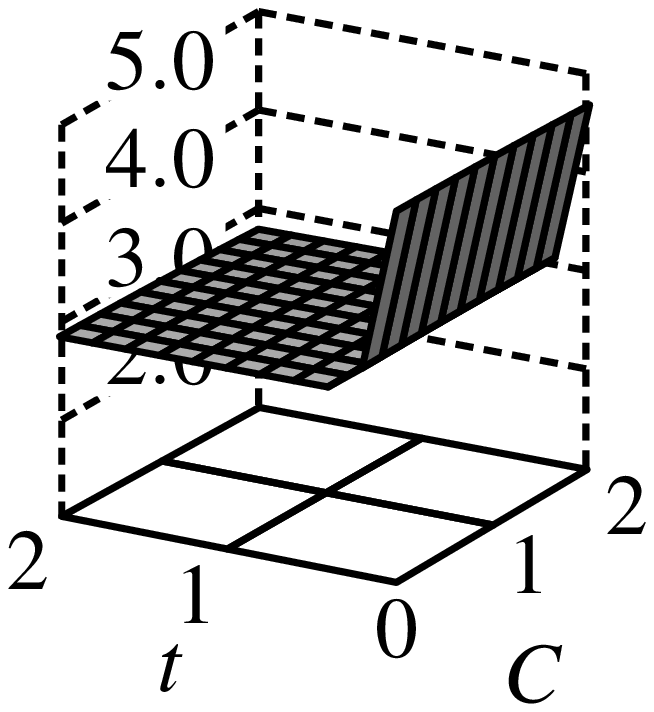}}
  \\
  \subfloat[$\mu=1$,$\nu=1$]{\label{subfig:wsurfthree}
    \includegraphics[width=58mm]{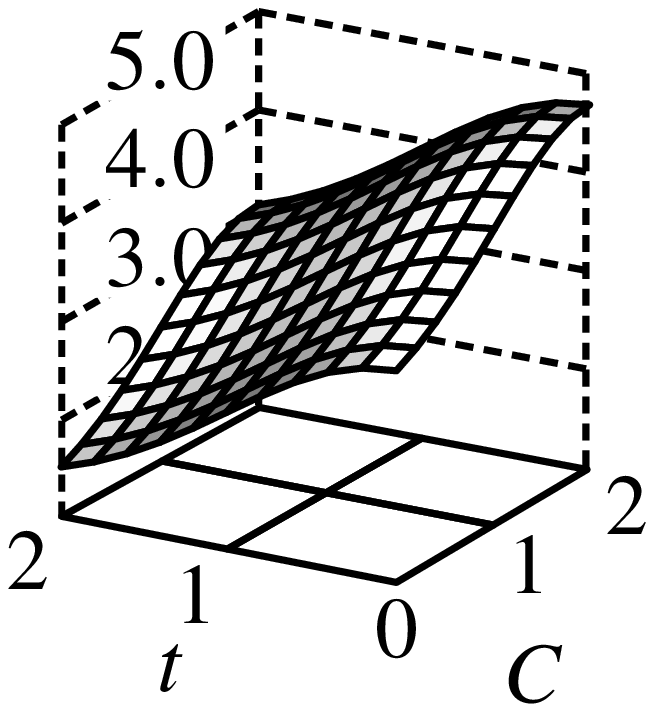}}
  \subfloat[$\mu=1$,$\nu=0.1$]{\label{subfig:wsurffour}
    \includegraphics[width=58mm]{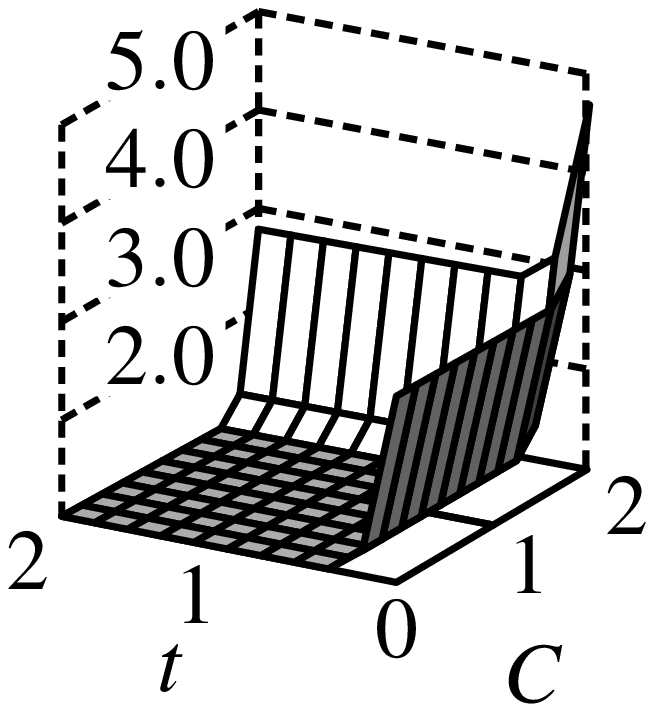}}
  \caption[Take your pants upstairs]{Fitness surfaces for males
    bearing ornaments: fitness slopes away from its maximum at $t = 0$
    as the trait grows, and increases with more condition alleles.
    Decreasing $\mu$ decreases fitness for males with fewer than $2$
    condition alleles.  When selection on condition weakens ($\mu =
    10$, \subref{subfig:wsurfone} and \subref{subfig:wsurftwo}) the
    fitness profile flattens in the $C$-dimension; compare this to
    $\mu = 1$ (\subref{subfig:wsurfthree} and
    \subref{subfig:wsurffour}) where males with high-viability alleles
    at both condition loci have significantly higher fitness than
    those with $1$ or none.  Direct selection on the trait ($\nu$)
    follows a similar profile except that $\nu \leq 1$ such that only
    trait-less males have significantly higher fitness than
    trait-bearing males in all simulations.}
  \label{fig:wsurf}
\end{figure*}

Summed Gaussian functions of trait and condition describe viability
for males and females:
\begin{subequations}
  \begin{align}
    w_{m} (C,t) = & 1 +
    \exp\left(-\frac{(\mathcal{C}-C)^{2}}{2\mu^{2}}\right) +
    \exp\left(-\frac{t^{2}}{2\nu^{2}}\right) \\
    w_{f} (C) = & \left(1 +
      \exp\left(-\frac{(\mathcal{C}-C)^{2}}{2\mu^{2}}\right)\right)
  \end{align}
\end{subequations}
where males have $\mathcal{C}$ condition loci, $\mu$ sets the relative
strength of selection for condition, $\nu$ determines the relative
strength of selection against the trait.  Smaller values of $\mu$ and
$\nu$ correspond to stronger selection (see Figure~\ref{fig:wsurf}).

The frequency of haplotypes changes through viability selection:
\begin{equation}\label{eq:viability}
  P_{i}^{\prime} (y) = \frac{P_{i} (y)W_{i} (y)}{\bar{W} (y)} 
\end{equation}
where $P_i (y)$ and $P_{i}^{\prime} (y)$ represent the frequency of
haplotype $i$ at age $y$ before and after selection, respectively.
$W_{i} (y)$ correspondingly represents the viability of haplotype $i$
at age $y$ and $\bar{W} (y)$ represents mean viability within age
class $y$.  The matrix $\mathbf{M}$ expresses the probability of
mating between a female of genotype $i$ and a male of genotype $j$:
\begin{equation}
  \label{eq:mtable}
  \mathbf{M}_{ij} = \frac{P_{i}^{\prime}\sum_{y = 0}^{y_{max}} \phi_{i} \left(t_{j}
      (y)\right)P_{j}^{\prime}(y)\pi(y)} 
  {\sum_{l}\sum_{\upsilon=0}^{y_{max}} \phi_{i} \left(t_{l}
      (\upsilon)\right)P_{l}^{\prime} (\upsilon)\pi (\upsilon)}
\end{equation}
where $\pi(y)$ represents the frequency of males of age $y$, and $t_j
(y)$ represents the trait size of a male of genotype $j$ at age $y$.
Condition is not an argument to the $t ()$ function in this case, as
it was in Equation~(\ref{eq:btrait}) since the genotype $j$ specifies
the male's condition.  Equation~(\ref{eq:mtable}) expresses the
frequency of matings by summing over male ages the product of female
mating rate ($\phi$, a function of trait size) and the frequency of
the male haplotype $j$ in the general population, i.e.\ adjusted for
the age structure $\pi$.  The subscript $i$ on mating rate refers to
choosy versus non-choosy females.  The denominator adds up the
age-dependent sums across all male haplotypes.  The ratio of these
summed terms then yields the probability that a male breeding with a
female of haplotype $i$ is of haplotype $j$.  Multiplying by the
female haplotype $i$ frequency after selection $P_{i}^{\prime}$ yields
the frequency of $ij$ pairs after mate choice.

Summing the product of mating probabilities and recombination
probabilities across $\mathbf{M}$ yields the frequency of new zygotes
\begin{equation}
  P_{i}^{\prime} (0) = \sum_{jk}R_{jk\rightarrow i}M_{jk}
\end{equation}
where $R_{jk\rightarrow i}$ represents the proportion of $jk$ matings
that yield genotype $i$ after recombination
\citep{burger00:_mathem_theor_selec_recom_mutat}.  Condition loci
recombine freely ($r = 0.5$) with each other and other loci; other
loci recombine at arbitrary frequencies ($0\leq r \leq 0.5$; see
Table~\ref{tab:params}).  Condition loci recombine freely so that they
will represent unlinked loci far away in the genome, and could also
represent multiple unlinked loci \citep{rowe1996lpa}.  The trait and
preference loci recombine at arbitrary frequency since prior works
show that recombination frequency affects indirect selection on
preference \citep{kirkpatrick1997strength,KP:82}.

The relative size of the zygote class is found by the age-weighted sum
of the mean fecundities of all adult age classes:
\begin{equation}
  \label{eq:zygsize}
  \pi^{\prime} (0) = \sum_{y=0}^{y_{max}}\bar{m} (y) \pi (y)
\end{equation}
where $m (y)$ is the fecundity of an individual aged $y$.  The new
relative size of an adult age class is given by the mean viability of
the age class:
\begin{align}
  \label{eq:adultsize}
  \pi^{\prime} (y) = & \bar{W} (y) \pi (y) \\
  = & \pi (y)\sum_{i} P_{i} (y) W_{i} (y).
\end{align}
I then calculate the new age distribution by dividing by the sum of
all new age class sizes \citep{MooradAugust2008}:
\begin{equation}
  \label{eq:moorad}
  \pi^{\prime\prime}(y) = \frac{\pi^{\prime} (y)}{\sum_{{\upsilon}= 0}^{y_{max}}\pi^{{}\prime} ({\upsilon})}.
\end{equation}

I calculated the initial age structure by specifying $\lambda$, the
geometric rate of increase for a population in stable age
distribution.  I then used this Gaussian survivorship function
centered at $0$ to calculate survival probabilities:
\begin{equation}
  \label{eq:lofx}
  l (y) = \exp \left(\frac{-y^{2}}{2}\right). 
\end{equation}
The age distribution is then given by \citep[see][]{CH:94}:
\begin{equation}
  \label{eq:sad}
  \pi (y) = \frac{\lambda^{-y}l (y)}
  {\sum_{\upsilon=0}^{y_{max}}\lambda^{-\upsilon}l (\upsilon)}.
\end{equation}

All simulations started with the trait allele $\ttwo$ and preference
allele $\ptwo$ at non-zero frequencies in the youngest age-class, and
zero in the older age-classes.  Simulations ran until:
\begin{inparaenum}[(1)]
\item the trait allele $\ttwo$ fixed;
\item the trait allele $\ttwo$ or preference allele $\ptwo$ was lost
  (frequency dropped below $10^{-12}$);
\item the preference allele $\ptwo$ fixed;
\item the Euclidean distance between successive generations in
  preference and trait allele frequencies dropped below $10^{-9}$; or
\item the simulation ran for $10$ million iterations.
\end{inparaenum}

\section{Results}

\begin{table*}
  \begin{centering}
    \begin{tabular}{lll}
      Symbol & Meaning & Values\\\hline
      $t$ & Male trait size &$0-27.30$\\
      $\alpha$ & Coefficient of preference &$0.6$,$1$ \\
      $\mu$ & Coefficient of selection on condition & $1$,$10$\\
      $\nu$ & Coefficient of selection on trait & $0.0001-1.0$ in steps
      of $0.02$ \\
      $r$ & Recombination fraction & $0.25,0.5$ \\
      $\lambda$ & Geometric rate of increase (initial age structure) & $1.0$\\
      $b$ & Coefficient of trait growth & $0.0-0.5$ in steps
      of $0.02$\\
      $\mathcal{C}$ &Number of condition loci & $2$ \\
      $y_{max}$ & Oldest age for males & $2$\\
      $y_{max}+1$ & Number of age classes & $3$ 
    \end{tabular}
    \caption{List of variables and parameters with typical
      values \label{tab:params}}
  \end{centering}
\end{table*}

\subsection{Selection on male trait and growth}
\label{sec:bynu}
\begin{figure*}
  \begin{center}
    \subfloat[Age-dependence]
    {\label{subfig:bynuplane.5}
      \includegraphics[width=0.45\textwidth]{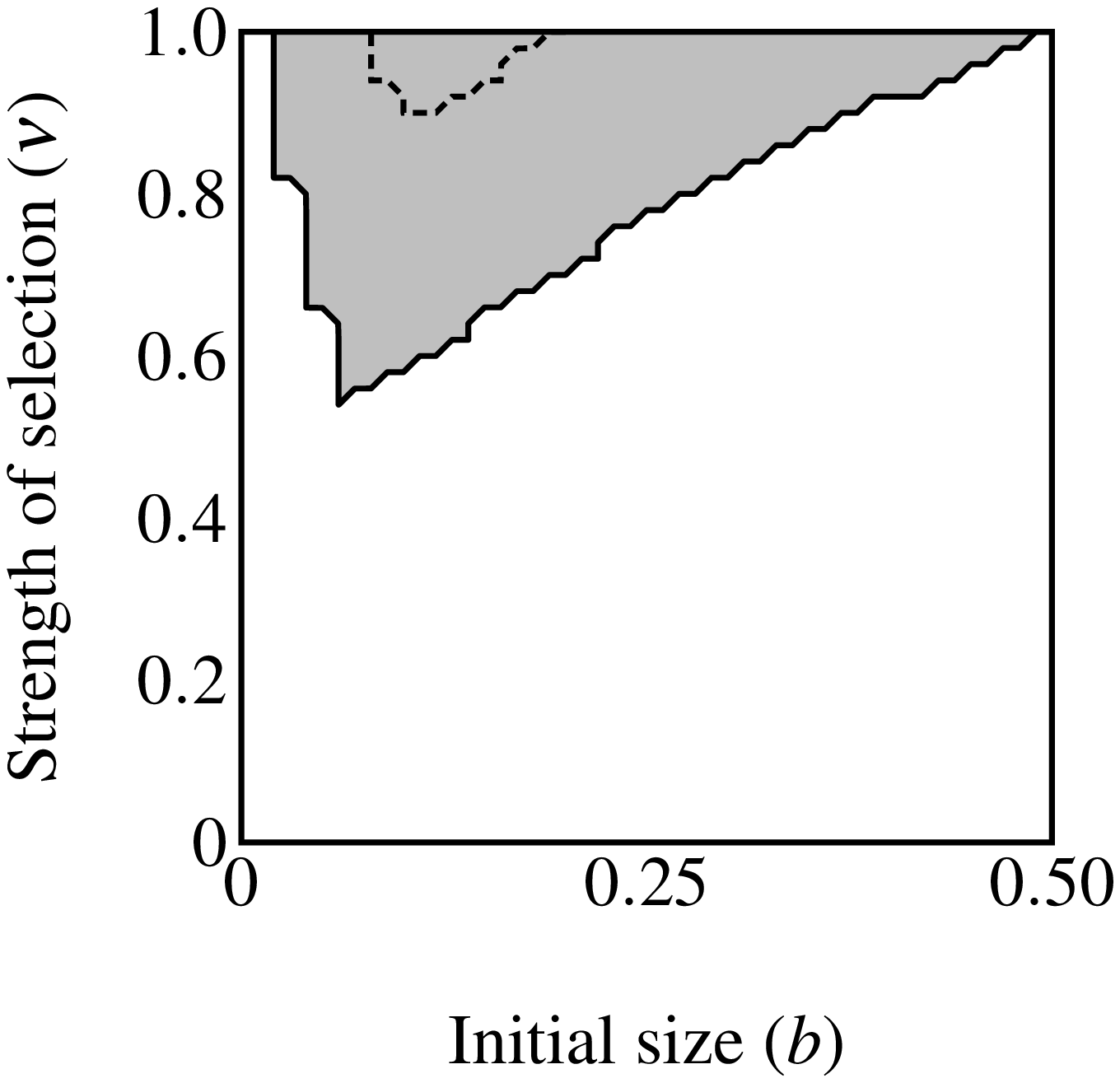}}
    \subfloat[Age-independence]
    {\label{subfig:bynuplane.max}
      \includegraphics[width=0.45\textwidth]{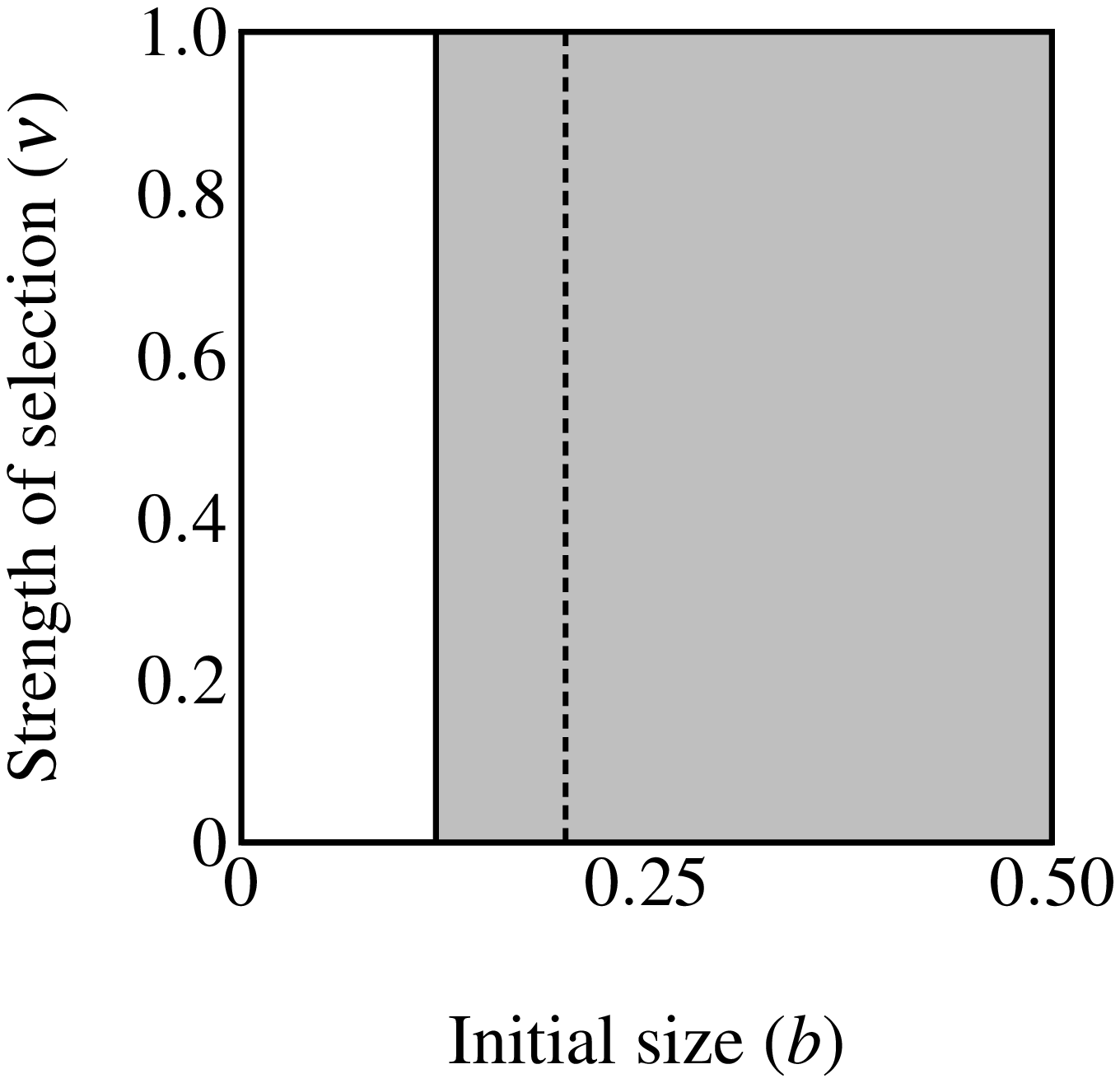}}
    \caption[Region of trait fixation for age-dependent and
    age-independent traits]{Region of trait fixation (light gray) for
      age-dependent \subref{subfig:bynuplane.5} and age-independent
      traits \subref{subfig:bynuplane.max} in a plane defined by
      strength of selection against the trait ($\nu$) and the growth
      coefficient of the trait ($b$).  Dashed line indicates region of
      fixation under $\alpha = 0.6$; solid lines indicates region of
      fixation for $\alpha = 1.0$.  The region of fixation for
      $\alpha=0.6$ is contained within the region of fixation for
      $\alpha=1.0$ in both panels.  Under age-dependence $b$ is the
      value of the trait for a $0$-year old male.  The $b$-axis
      corresponds linearly to the trait size of ornamented males ($t
      (C,y) = b e^{C y}$), such that populations depicted further to
      the right on the $b$-axis will have larger average trait sizes.
      Smaller values of $\nu$ correspond to stronger selection.}
    \label{fig:bynuplane}
  \end{center}
\end{figure*}

The first set of simulations sought to determine the important
parameters for evolution of the age-dependent trait compared with an
age-independent trait.  The age-independent trait was the same size as
the eventual size of the age-dependent trait.  Age-dependent
simulations ran with coefficient of preference ($\alpha$),
recombination frequency ($r$) and selection on condition ($\mu$) at
the values indicated in Table~\ref{tab:params}.  I used initial values
of $p_{C} = 0.01$, $\pp = 0.1$, $\pt = 0.001$, and $\pf = 0.0$.  I
compare this to a population where all initial values were the same
except that males were fixed for $\ftwo$, i.e.\ male trait expression
was age-independent and expressed at the largest size attainable ($t =
b \exp(C y_{max})$).  I analyzed the relative roles of selection
intensity and trait size by plotting the equilibrium value of $\pt$
(fixation versus loss of the trait allele) over a plane defined by
$0.00001 \leq \nu \leq 1.0$ and $0 \leq b \leq 0.5$ (see
Table~\ref{tab:params}; see Figures~\ref{fig:bynuplane} and
~\ref{fig:bynuplane.poly}).

The area of parameter space where the trait fixes depends on three
parameters: $\alpha$ (strength of preference), $b$ (``growth
coefficient''; see Equation~(\ref{eq:btrait})) and $\nu$ (strength of
selection; Figure~\ref{fig:bynuplane}).  Recombination frequency ($r$)
and selection on condition ($\mu$) do not appear to qualitatively
affect the results in these simulations.  Strength of preference
affects the ``readiness to mate'' of females over the range of trait
values present in these simulations: when $\alpha = 0.6$ a choosy
female is $9.20$ times more likely to mate with a $2$ year-old,
high-condition male with a growth coefficient of $b = 0.25$.  At the
higher value of $\alpha = 1.0$, a choosy female is $14.65$ times more
likely to mate with the same trait-bearing male.  We see a qualitative
difference between populations with age-dependent traits and
populations with age-independent traits at both values of $\alpha$.
The selection parameter $\nu$ determines the pattern of fixation for
age-dependent traits, but has no effect on age-independent traits.  At
$\alpha = 0.6$ the trait fixes in a very small portion of the $b-\nu$
plane near $\nu = 1.0$ in age-dependent simulations.  The traits fixes
above the threshold $b$ value of $0.20$ in the corresponding
age-independent simulations.  At $\alpha = 1.0$ the regions of
fixation are larger, but the qualitative differences between
age-dependent and age-independent expression remain: the age-dependent
trait fixes in a roughly triangular region characterized by relatively
weak selection and containing a region of small initial trait sizes
(Figure~\ref{subfig:bynuplane.5}).  This contrasts to age-independent
simulations, where again above a threshold size ($b = 0.12$) the trait
fixes independently of selection intensity
(Figure~\ref{subfig:bynuplane.max}).

\begin{figure*}
  \centering
  \subfloat[$b = 0.1$,$\nu = 0.8$]
  {\label{subfig:traj.1.8}
    \includegraphics[width=0.45\textwidth]{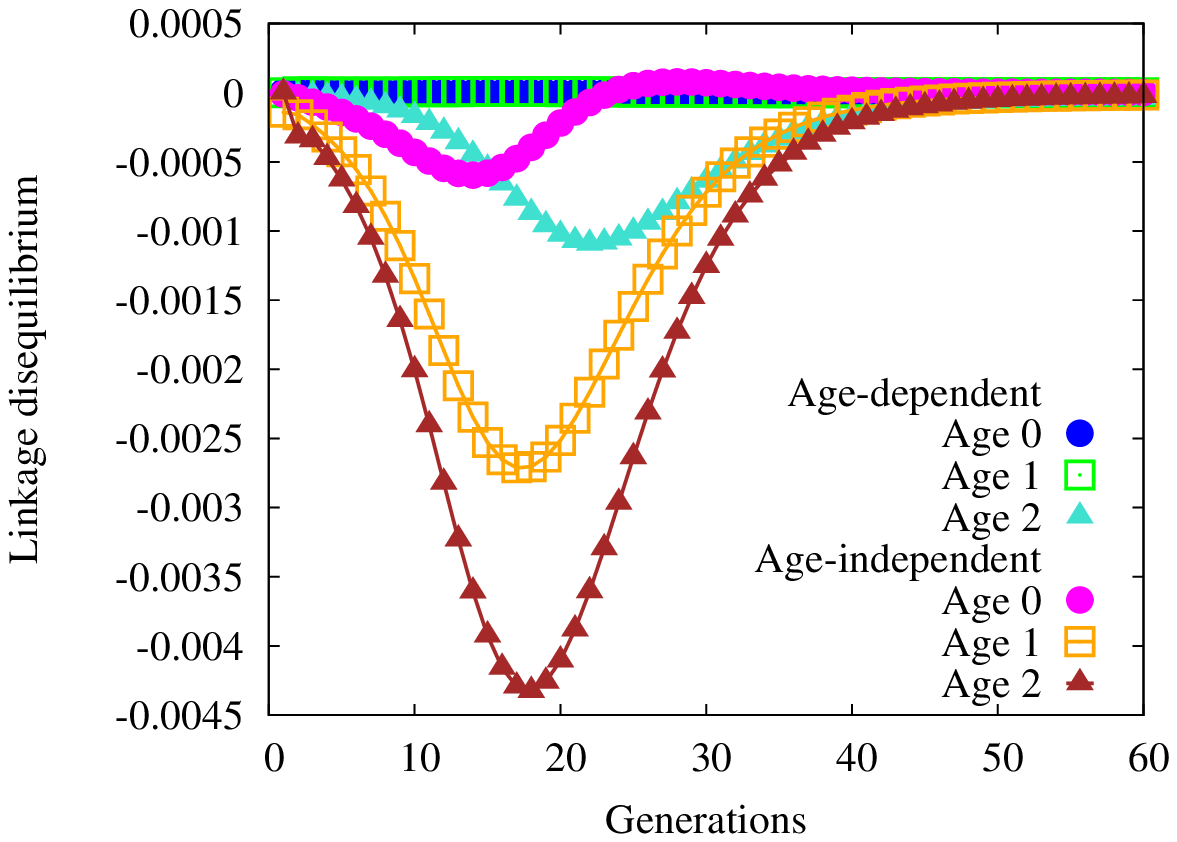}}
  \subfloat[$b = 0.2$,$\nu = 0.8$]
  {\label{subfig:traj.2.8}
    \includegraphics[width=0.45\textwidth]{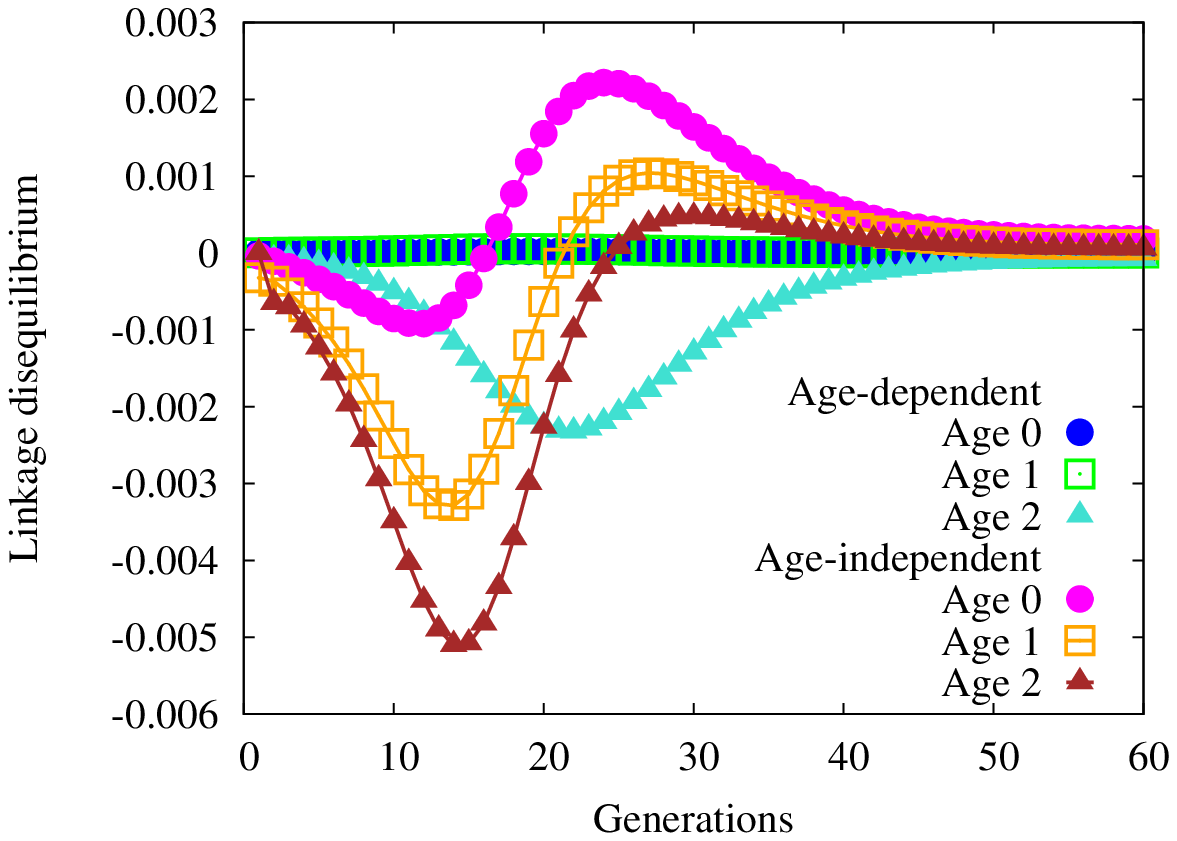}}\\
  \subfloat[$b = 0.1$,$\nu = 0.2$]
  {\label{subfig:traj.1.2}
    \includegraphics[width=0.45\textwidth]{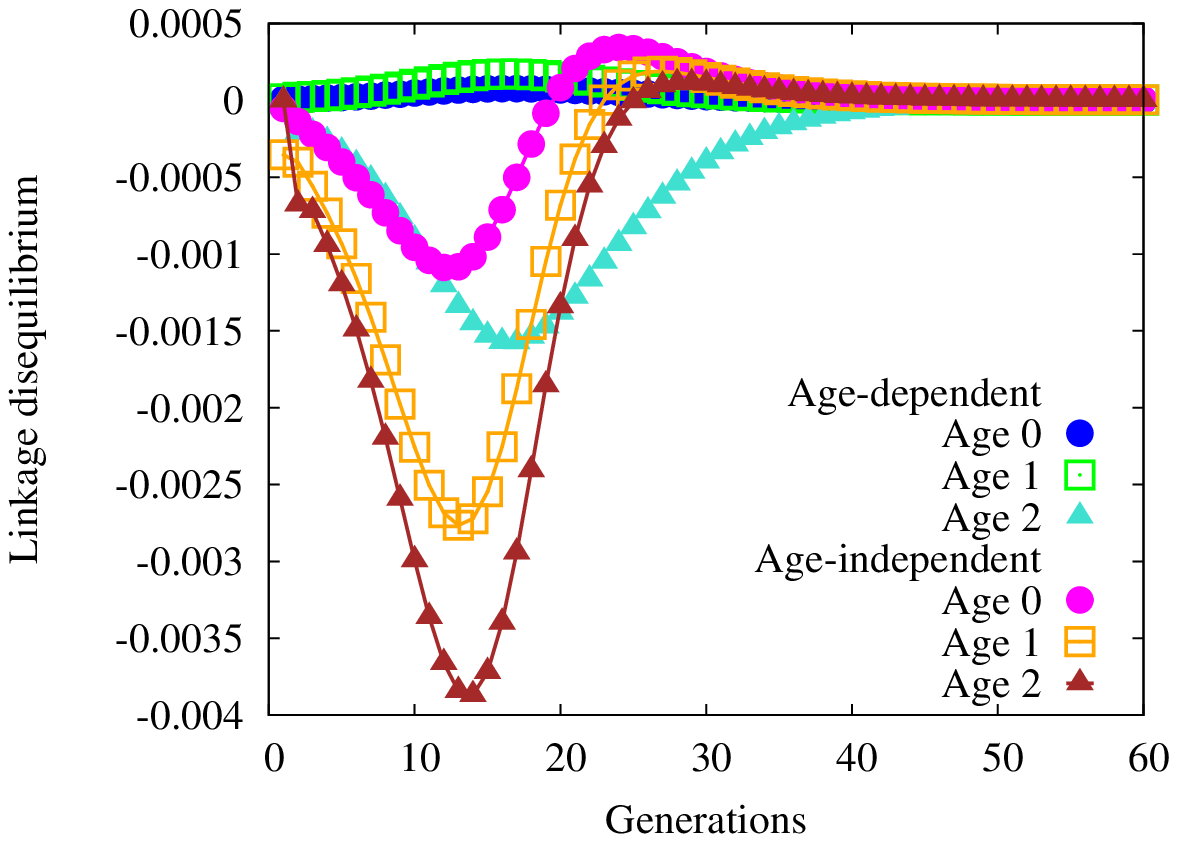}}
  \subfloat[$b = 0.2$,$\nu = 0.2$]
  {\label{subfig:traj.2.2}
    \includegraphics[width=0.45\textwidth]{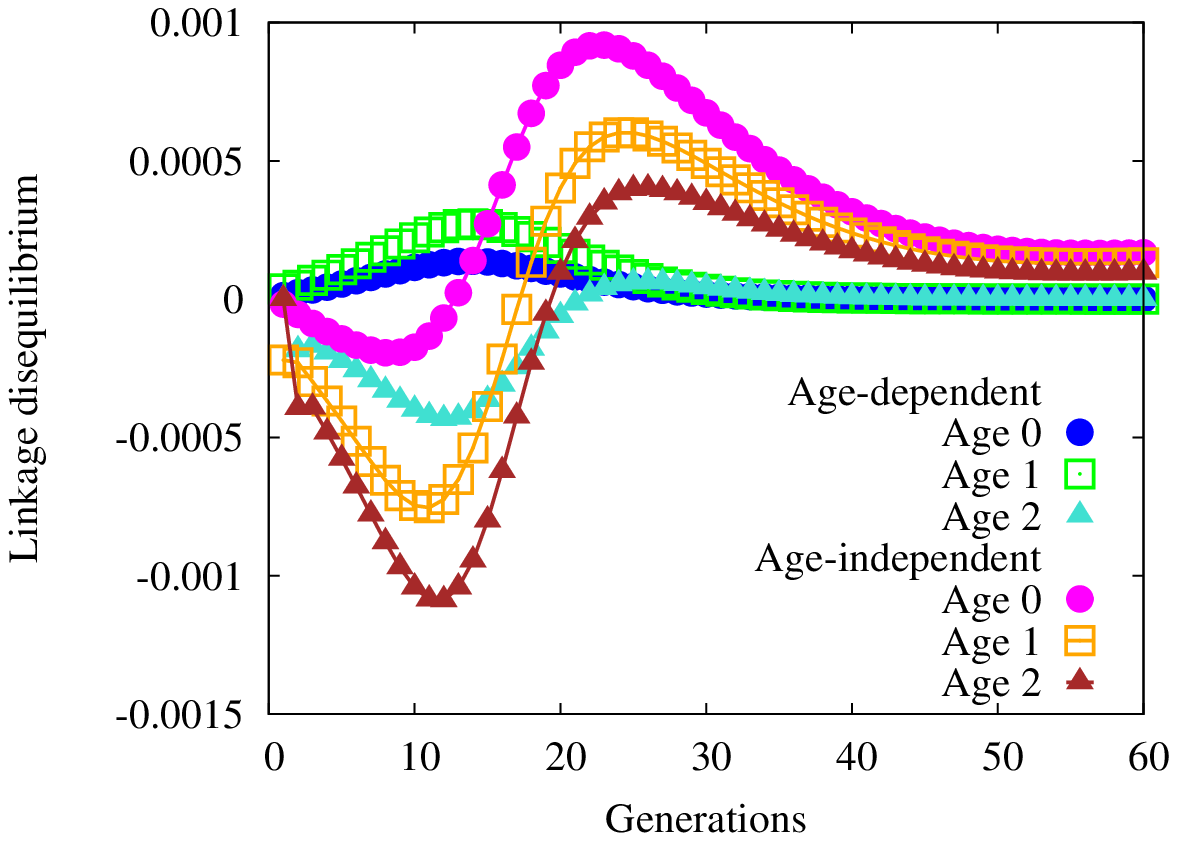}}
  \caption[Trajectories of linkage disequilibrium between the trait
  and one of the condition loci]{Trajectories of linkage
    disequilibrium between the trait and one of the condition loci,
    with $\alpha = 1.0$.  This quantity reflects selection on the
    genotype bearing both the trait allele and the beneficial
    condition allele.  The four panels are defined by regions of
    Figure~\ref{fig:bynuplane} with different patterns of fixation:
    \subref{subfig:traj.1.8} fixation under age-dependence, not under
    age-independence; \subref{subfig:traj.2.8} fixation under both
    modes of expression; \subref{subfig:traj.1.2} no fixation under
    age-dependence or age-independence; and \subref{subfig:traj.2.2}
    fixation only under age-independence.  Note the differing scales
    on the vertical axis of each panel.  Quantities not adjusted for
    age structure. }
\label{fig:trajectories}
\end{figure*}

Figure~\ref{fig:trajectories} shows a sample of linkage disequilibrium
trajectories from the four regions defined by fixation and loss of
age-dependent and age-independent traits.
Each point on a curve represents the statistical association (linkage
disequilibrium) between a beneficial condition allele and the ornament
allele.  A positive association indicates that the trait allele occurs
in the same genotype with a beneficial condition allele more often
than expected by chance.  A negative association indicates that the
two alleles are found less often in one genotype than expected by
chance.  When the alleles randomly associate e.g.\ at the beginning of
the simulation, or when one or both alleles are fixed or lost, the
linkage disequilibrium equals zero.  Condition alleles do not fix in
these simulations due to biased mutation.

%
These associations indicate the effect of selection on genotypes that
act as indicators of condition, i.e.\ have both a condition allele and
a trait allele.  When selection favors condition (always) and does not
favor the trait, it will drive alleles apart so that they are rarely
found in the same genotype, leading to the negative associations found
in Figure~\ref{fig:trajectories}.  I caution the reader against
interpreting the values in Figure~\ref{fig:trajectories} as directly
indicative of the strength of overall selection, since the quantities
do not include adjustment for age structure.  Each point reflects the
action of selection on a particular cohort, and not the strength of
selection against the trait in the population.  For example, the
larger absolute values in older age-classes in the figure does not
indicate that selection acts more strongly on individuals in that age
class.  Selection works quite weakly on older age classes, since old
individuals are quite rare.

Using these trajectories gives us a slightly clearer picture of the
forces acting to generate the pattern of fixation and loss seen in
Figure~\ref{fig:bynuplane}.  Each panel shows curves for each of three
age classes under both age-dependent and age-independent simulations.
All four panels show stronger associations in age-independent
simulations than in age-dependent simulations.  We can also see that
magnitude (absolute value) increases from age $0$ to age $2$.
Increasing negative magnitude reflects that high-condition males,
likely to survive to older age classes, are less likely to be
ornamented.  However, this effect is generally much weaker in
age-dependent populations than in age-independent populations
(Figure~\ref{subfig:traj.1.8}).  Males that reach the oldest age class
under age-dependence are more likely to be high-condition and
ornamented than under age-independence.  More raw material for sexual
selection remains within a cohort under age-dependence.

The right-hand column of Figure~\ref{fig:trajectories} shows
trajectories for a larger trait ($b = 0.2$).  A similar curve appears
for age-dependent traits in Figure~\ref{subfig:traj.2.8} as in
Figure~\ref{subfig:traj.1.8}.  However, the age-independent trait
responds to selection rapidly, showing a different shape from
age-dependent populations (Figure~\ref{subfig:traj.1.2} and
Figure~\ref{subfig:traj.2.2}).  Readers should keep in mind when
viewing this figure that the age structure of the population,
especially under strong selection, biases heavily toward young males.
Although linkage disequilibrium increases in older age classes, the
associations for age class $0$ most closely reflect the overall
linkage disequilibrium in the population as a whole.

\subsection{Mode of development}

Another set of simulations sought to determine the crucial parameters
favoring age-dependent expression over age-independent expression.
These simulations began with polymorphism at the $\flocus$ locus,
i.e. $\ftwo$ began at a low, non-zero frequency.  Fixation of $\ftwo$
depended on the level of trait expression chosen for the
age-independent males in that simulation, one of
\begin{inparaenum}[(1)]
\item $b$;
\item $\bar{t}$; or
\item $t_{max} = b \exp(C y_{max})$.
\end{inparaenum}
I simulated these conditions with initial values of $p_{C} = 0.01$,
$\pp = 0.1$, $\pt = 0.1$, and $\pf = 0.1$ in all age classes.
Age-independent males occurred with one particular trait function ($t
= b$, $t = \bar{t}$, or $t = b\exp(C y_{max})$) for a specific
simulation: when the $\ftwo$ allele caused males to have smaller
traits than the oldest age-dependent males then $\ftwo$ was lost
everywhere in the $b$-$\nu$-plane.

\begin{figure*}
  \begin{center}
    \subfloat[$\mu = 1.0$]
    {\label{subfig:bynuplane.poly.5}
      \includegraphics[width=0.45\textwidth]{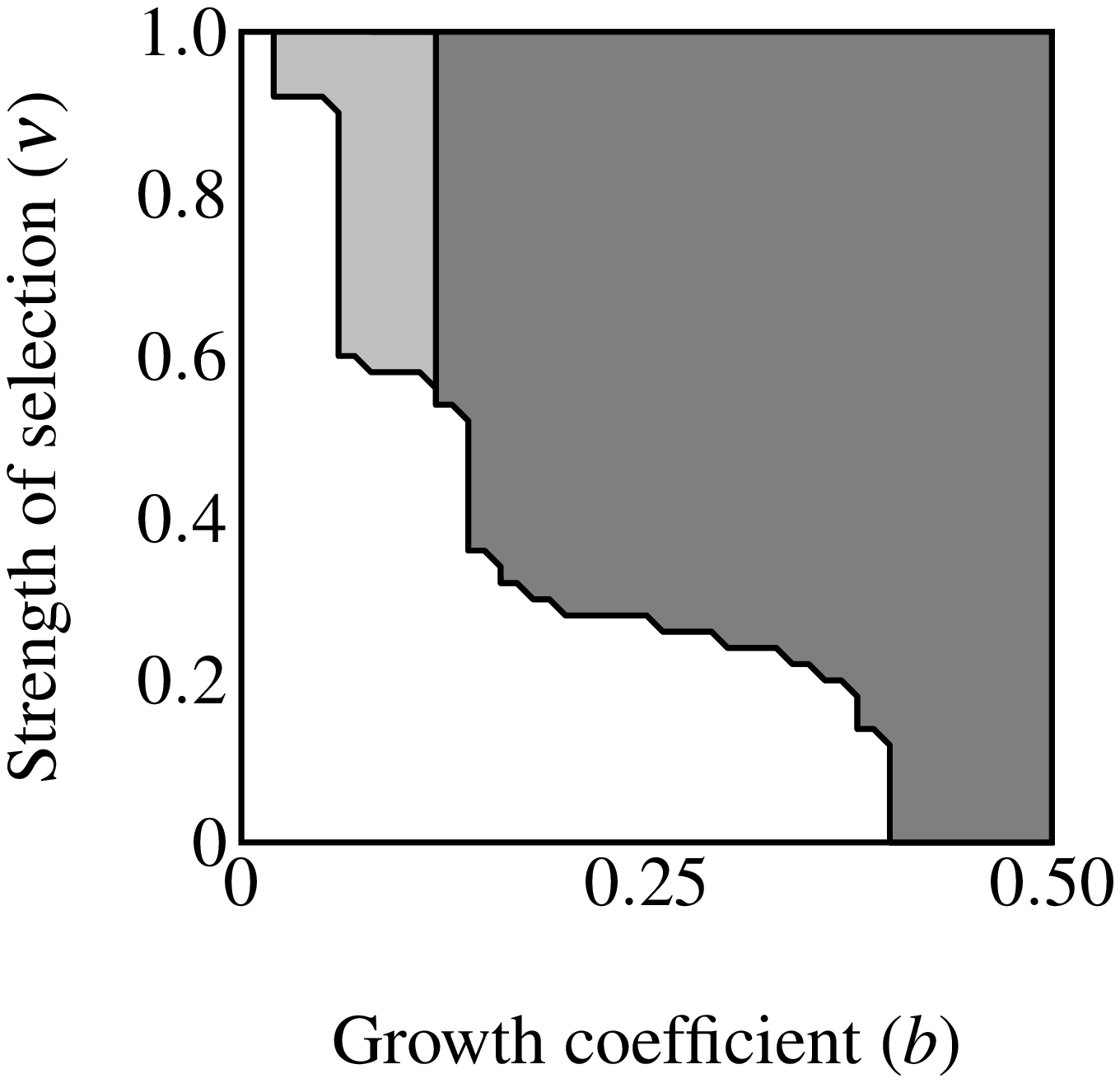}}
    \subfloat[$\mu = 10.0$]
    {\label{subfig:bynuplane.poly.mu}
      \includegraphics[width=0.45\textwidth]{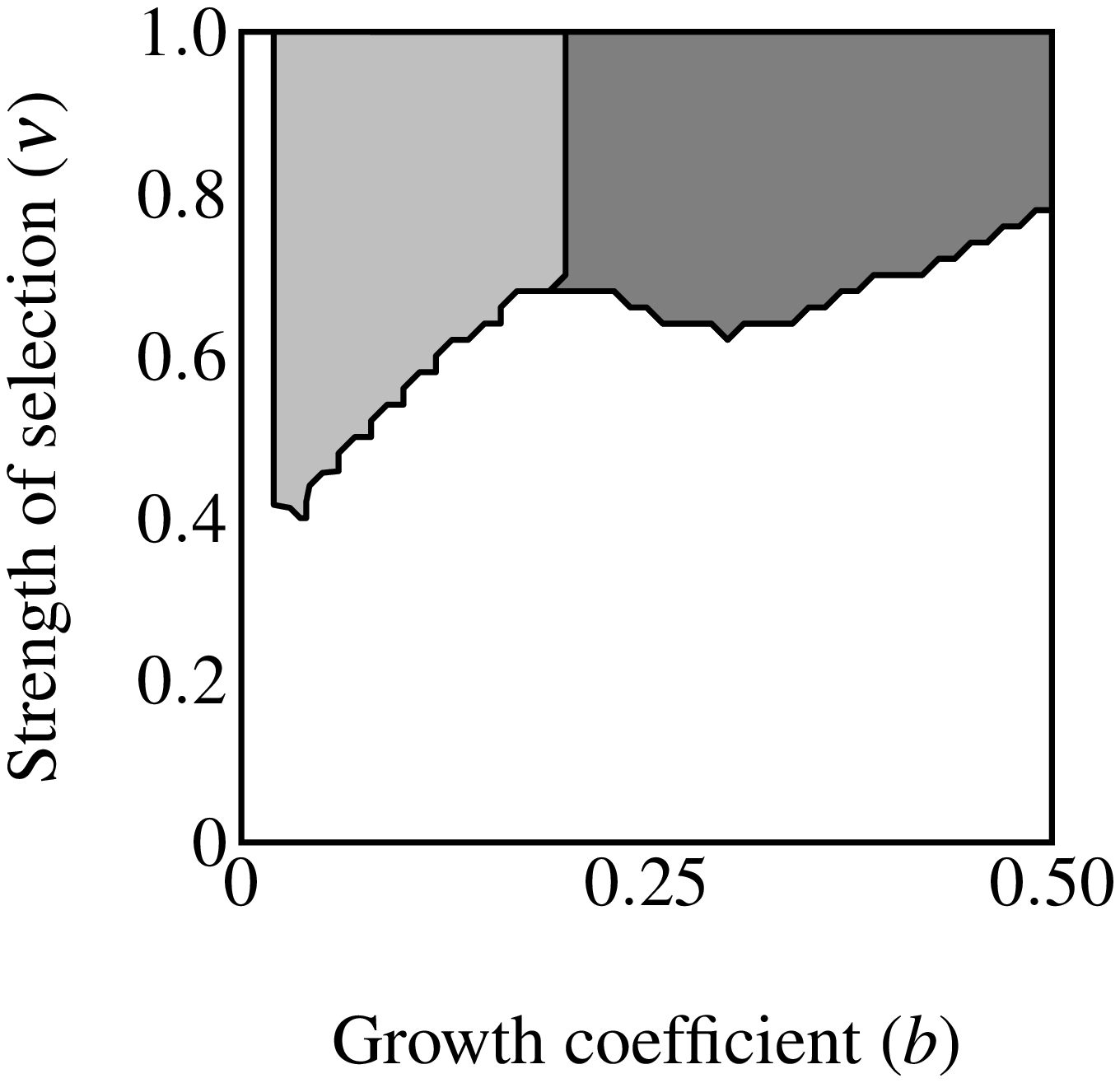}}
    \caption[Regions of fixation for the trait only and the trait
    along with age-independent mode of development allele]{Regions of
      fixation for the trait only (light gray) and the trait along
      with age-independent mode of development allele ($\ftwo$; dark
      gray).  Panel \subref{subfig:bynuplane.poly.5} shows strong
      selection on condition and panel
      \subref{subfig:bynuplane.poly.mu} shows weak selection on
      condition.  The trait is lost in the white region.  $\ftwo$ does
      not change from its initial frequency of $0.1$ in the light gray
      region, resulting in a predominance of age-dependent signaling
      at small trait sizes.}
    \label{fig:bynuplane.poly}
  \end{center}
\end{figure*}

When males carrying $\ftwo$ expressed the trait range of the oldest
age-dependent males throughout their lives ($t = b\exp(C y_{max})$)
the $\ftwo$ allele fixed in an area including the highest intensity of
selection (Figure~\ref{fig:bynuplane.poly}).  However, the $\ftwo$
allele also failed to increase in an area of small initial trait
values where the trait did fix.  In other words, expression remained
polymorphic in this region, with the majority of males displaying
age-dependence (light gray in Figure~\ref{subfig:bynuplane.poly.5}).
This effect is even stronger when selection on condition is relaxed
($\mu = 10$; see Figure~\ref{subfig:bynuplane.poly.mu}).  The area of
trait fixation where $\flocus$ remained polymorphic is larger when
$\mu = 10$ and includes regions of more intense selection against the
trait as well as requiring larger trait sizes for fixation of $\ftwo$.
Age-dependence occurs at larger trait sizes and under more intense
selection when the population displays more variance in condition.
This suggests that for a given $b$, age-independence only brings males
a mating advantage with strong selection on condition.
%
%
%

I also repeated the above simulations with $\pf = 0.9$ initially.  For
all parameter values $\ttwo$ was lost below a certain threshold value
and above this threshold, both $\ttwo$ and $\ftwo$ fixed.  This
reinforces the above results: when the majority of males display
age-independent expression, trait size alone determines the fixation
of the trait.

\begin{figure*}
  \begin{center}
    \subfloat[$\mu = 1.0$]
    {\label{subfig:bynuplane.poly.fixed.5}
      \includegraphics[width=0.45\textwidth]{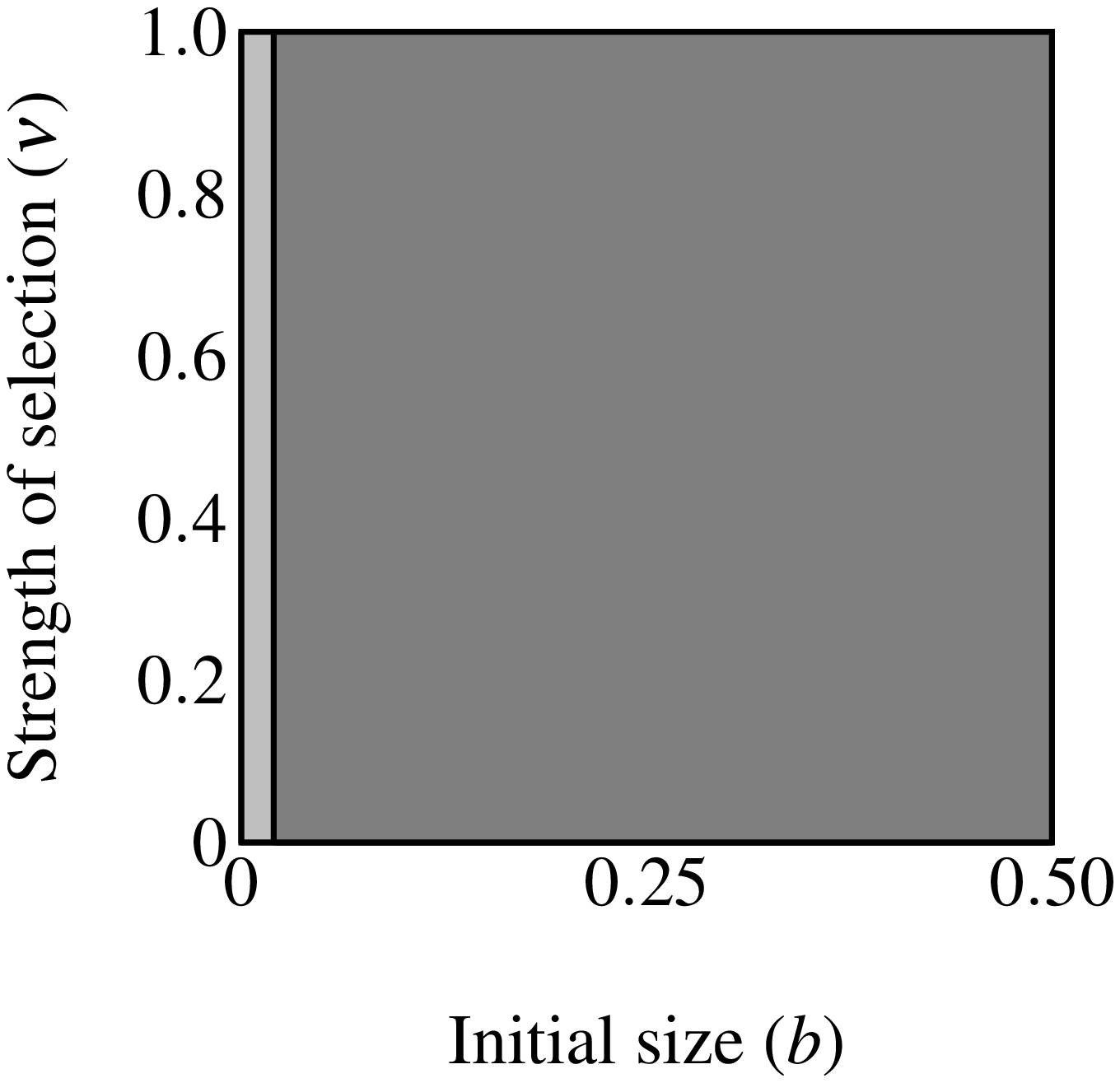}}
    \subfloat[$\mu = 10.0$]
    {\label{subfig:bynuplane.poly.fixed.mu}
      \includegraphics[width=0.45\textwidth]{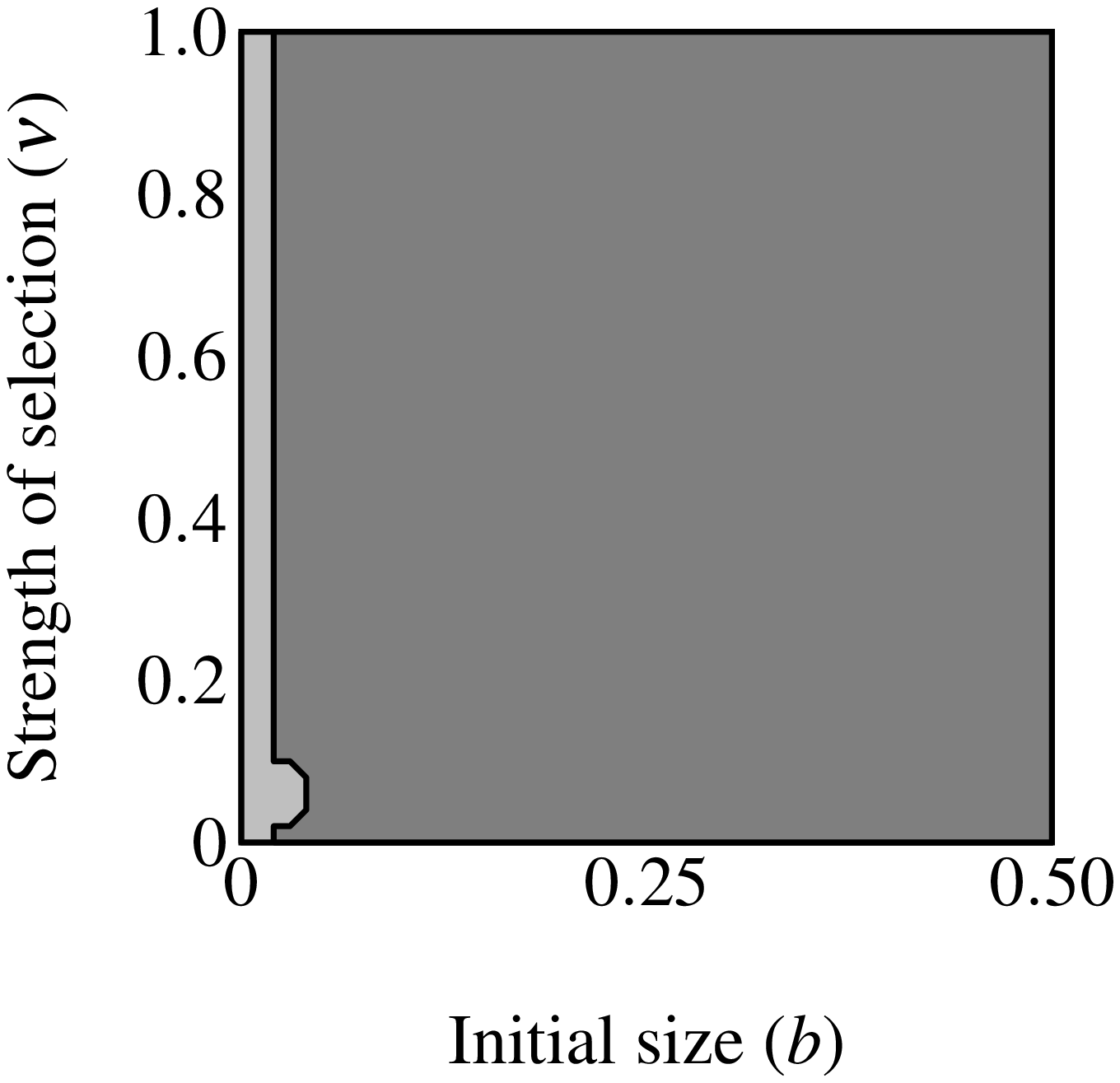}}
    \caption[Regions of fixation for the age-dependent trait and the
    age-independent trait with the trait and preference initially
    fixed]{Regions of fixation for the age-dependent trait (light
      gray) and the age-independent trait ($\ftwo$; dark gray) with
      the trait and preference initially fixed.  Panel
      \subref{subfig:bynuplane.poly.fixed.5} shows strong selection on
      condition and panel \subref{subfig:bynuplane.poly.fixed.mu}
      shows weak selection on condition.  The $\ftwo$ allele supplants
      age-dependent signaling at any strength of selection above a
      certain threshold.  Age-dependent signaling persists at small
      trait sizes.}
    \label{fig:bynuplane.poly.fixed}
  \end{center}
\end{figure*}

Another set of simulations considered the fixation of the $\ftwo$
allele with established traits ($\pt = 1.0$) and preferences ($\pp =
1.0$; Figure~\ref{fig:bynuplane.poly.fixed}).  The $\ftwo$ allele was
initially rare ($\pf = 0.1$).  Individuals carrying $\ftwo$ displayed
the maximum trait size of age-independent males of comparable
condition at their maximum age ($t = b\exp(C y_{max})$).
Age-independent signaling came to predominate regardless of selection
strength above a certain small size threshold.  Below this threshold
we can say that age-dependent signaling persisted.  The same pattern
was observed with a minor exception under strong selection against the
trait, regardless of selection on condition
(Figure~\ref{subfig:bynuplane.poly.fixed.mu}): selection on condition
does not appear to qualitatively affect the size at which
age-independent signaling out-competes age-dependent signaling.

\section{Discussion}
\label{sec:discussion}
Although age-dependent sexual traits are common in nature, their
evolution presents several dynamical difficulties.  Every sexually
selected trait balances the benefits derived through mate choice with
the costs of growing a trait.  Age-dependent traits undergo a period
when selection can easily eliminate them.  I have shown that an
age-dependent trait can evolve under relatively weak selection at a
wider range of trait sizes than can an age-independent trait
(Figure~\ref{fig:bynuplane}).  While an age-dependent trait can evolve
by producing small traits at young ages when selection is intense,
larger initial trait sizes can only occur under proportionally less
intense selection.  This contrasts strongly with the age-independent
simulations.  Size of the trait determines the eventual fixation of
the age-independent trait allele, regardless of selection intensity.
Simulations with genetic variation for mode of development
(age-dependent vs. age-independent) show that age-independence only
evolves at larger trait sizes and age-dependence can predominate under
weaker selection at smaller trait sizes.  Invasion of age-independent
traits occurred at relatively smaller trait sizes than when increasing
from rarity.  Again this only depended on trait size.  Evolution of
age-independence generally depends on trait size and not on the
intensity of selection against the trait.

Altogether the simulations show that small traits in young males
``solve'' the dynamical problem of reduced heritability.  The results
suggest that three factors favor the evolution of age-dependent sexual
signals:
\begin{inparaenum}[(1)]
\item low adult male mortality;
\item weak selection against the trait; and
\item strong age-dependence, i.e.\ small initial trait sizes.
\end{inparaenum}
This result makes sense from a life-history strategy perspective.
When adult mortality is lower, young males will invest less in
reproductive strategies, producing a negative correlation between
expected future reproduction and age-specific investment.  Males can
afford to invest in mate attraction when adult mortality is low.
Age-dependence effectively partitions the life-history of males into
two stages dominated by two different fitness components.  Viability
takes precedence at young ages and mating success becomes more
important later in life.  This result parallels recent work on an
abundant primate showing that juvenile survival followed by mating
success form the most crucial fitness components
\citep{courtiol_natural_2012}.

Trajectories of linkage disequilibrium support the above theory
(Figure~\ref{fig:trajectories}).  Selection finds its greatest power
early in the life cycle, dealing with the majority of genetic
variation in the largest age class.  When the additive component of
that variation lies hidden within age-dependent traits, it can pass
through to older age classes where it will be fully expressed.  As
males age they reveal additive variance in condition, thus making them
more reliable signalers \citep[as in ][]{proulx2002oms}.  The
trajectories illustrate how selection can only eliminate the
age-dependent trait when either selection acts strongly or when the
young-age trait is large enough.  Weak selection (the top row of
Figure~\ref{fig:trajectories}) works even weaker on age-dependent
populations, so as to barely change allele frequencies until males get
older.  Age-dependence weakens selection against the trait.

The failure of the age-independent trait to evolve at lower trait
sizes (see Figure~\ref{subfig:bynuplane.max}) also supports the theory
presented above.  Using an individual perspective can aid
understanding of this result.  Consider a population early in the
evolution of the trait.  The trait and preference occur rarely, and
hence viability differences between ornamented and unornamented males
dominate the selection differential.  Mating success will not form a
significant part of fitness due to positive frequency dependence
(i.e.\ rarity of choosy females).  From an individual perspective, any
trait-bearing male will have lower fitness than an unornamented male.
For the trait to be advantageous, it must be large enough for males to
gain high mating success despite the rarity of choosy females.  On the
other hand, for trait-bearing males to avoid viability selection, it
must be small.  Compare young age-independent males to young
age-dependent males in this scenario.  Age-independent males that
avoid selection never become attractive.  Age-dependent males that
avoid viability selection, by contrast, do become more attractive.
Now consider males bearing large traits.  Age-independent males that
start out with sufficiently large traits immediately attract choosy
females, even though they probably die in the next round of selection.
If the reader picks a single $b$-value in Figure~\ref{fig:bynuplane},
this represent two potential males:
\begin{inparaenum}[(1)]
\item a fully-grown age-dependent male and
\item a young age-independent male.  
\end{inparaenum} Both males display the same trait size and have the
same viability costs.  However, the age-dependent male has already
lived through two mating episodes and three episodes of selection.
After mating both males will die, but the age-dependent male has
higher lifetime mating success.  Carrying attractive traits at a young
age reduces future mating opportunities.  We see a threshold of
attractiveness in Figure~\ref{subfig:bynuplane.max} where
age-independent males show large enough traits to avoid this cost.
\citet{kokko2001faH} has already noted that sexual advertisements are
life-history traits.  Considering the costs and benefits of
advertisements requires evaluation of the entire life cycle and all
the organism's interactions and requirements \citep{BA:02}.

The results here support the strategic modeling literature of
age-dependent signals
\citep{rands_dynamics_2011,proulx2002oms,kokko97:_evolut_stabl_strat_of_age}.
\citeauthor{proulx2002oms} modeled the situation where male longevity
and reproductive opportunities increase --- e.g.\ under a low adult
mortality environment --- and found that high-condition males downplay
their signaling relative to lower condition males, preserving
resources for survival.
\citeauthor{kokko97:_evolut_stabl_strat_of_age} came to the similar
conclusion that young males of lower condition should signal more than
their higher-condition cohorts, thus obscuring the observed
relationship between genetic quality and trait value.  Both studies
find that (optimally) males of a given condition signal inversely
proportional to number of remaining reproductive attempts (a predictor
of condition in any age-class).  Selection then favors females that
prefer to mate with older males, since they are more likely to be of
high condition.  These studies model competing strategies, whereas my
study uses a condition-dependent and age-dependent trait function to
model variation.  When selection weakens enough, with a particular
developmental trajectory, age-dependent signaling and female
preferences evolve in a population genetic model.  The evolutionary
dynamics, in this case, do mirror the conclusions of the optimization
models.  These results suggest that with the needed life-history
conditions and genetic variation we can expect selection on
life-histories to produce age-dependent signaling.


Empirical evidence and my results suggest that extending the male
lifespan facilitates sexual selection.  When traits are age-dependent
they can develop their most exaggerated forms at older ages when
selection is less intense.  Age-dependent traits or mating success
occur in a wide variety of taxa, including mammals
\citep{poissant08:_quant,pemberton04:_matin,clinton93:_sexual_selec_effec_male_life},
birds
\citep{hawkins_delayed_2012,evans_divergent_2011,taff_relationship_2011,Ballentine2009973,garamszegi_agedependent_2007,EVANS1997749},
fish
\citep{johnson_sexual_2011,jacob_male_2007,miller_effects_2005,candolin_changes_2000,candolin_increased_2000}
and insects
\citep{verburgt_male_2011,Judge2011185,kivleniece_sexual_2010,jones_elgar:2004,Jones2000}.
Despite the demographic problems of age-dependence, widespread
occurrence of age-dependent traits suggests that life-histories
promoting age-dependence are common.  Body size or traits directly
correlated with body size (e.g.\ weapons and ornaments) form the most
obvious example of age-dependent traits, and should satisfy the
assumptions of my model.  Researchers found age-dependent sexual
selection based on body size in Rocky Mountain Bighorn Sheep
(\emph{Ovis canadensis}; \citealt{Coltman2002}), who show a typical
life-history characterized by weakening viability selection and
increasing heritability over the lifespan.  Older males pay less of a
survival cost for larger bodies and larger horn sizes, facilitating
greater success in mating competition.  Certain behavioral and social
traits should display age-dependence under weak natural selection,
such as social network connectivity \citep{mcdonald_cooperative_1994}
and song repertoire \citep{Gil2001689}.  Age-based honesty also
creates an effective constraint producing age-dependence.
High-condition males cannot bypass age-dependence by ``faking'' the
trait.  Certain ``skills'' such as nest-building \citep{EVANS1997749}
show this form of honesty.


Readers should consider some limitations of the model I used here.  My
model inadequately portrays situations where direct costs of female
choice impact sexual selection.  Direct costs of choice could
considerably impact female survival and therefore a full analysis of
the life-history implications of sexual selection needs to include a
model of the female life-history.  Indirect costs, such as increasing
frequency of germ-line mutations with male age also greatly affect
female choice evolution in long-lived organisms
\citep{beck_evolution_2007,hansen1999aas}.  Female choice also depends
on female condition, which could produce strong negative linkage
disequilibrium (i.e.\ negative correlations) between condition loci
and preference loci under selection against the trait.
Condition-dependent female choice could therefore broaden the range of
parameters where an age-dependent trait can evolve.  The trait
function I used here also has somewhat narrow applicability, as it
strongly favors old males: age-dependent traits in nature probably
favor intermediate-aged males, who have become attractive but have not
deteriorated considerably \citep{Brooks2001308}.  A more
physiologically realistic trait function would peak in middle-age
\citep{johnson_sexual_2011}, but would tell us little more about the
evolution of the preference without considering costs of mating with
old (versus intermediate-aged) males.  I chose the small maximum age
of $2$ for computational efficiency.  Three age classes also yielded
young, middle- and old-aged males without producing old males with
enormously exaggerated traits.  When males had more age classes,
old-age traits became very large and far too advantageous to analyze
the tension between sexual selection and viability selection.  Males
in nature live in more complex age-structured populations with more
than three age-classes.  My model also uses an effectively infinite
population size and overestimates the effect of old males in the
population.  The effective population size of old age-classes could
diminish or fluctuate and drift could eliminate alleles for indicator
traits.

Age-dependent signaling offers a testable hypothesis relating
life-histories and sexual selection.  Researchers should check
independent developments of iteroparity and reduced adult mortality
for association with age-dependent sexual signals.  The weakening of
selection associated with lifespan development should facilitate
sexual selection by concomitant reductions in selection against
outrageous traits at older ages.  Further modeling should ask whether
senescence and accumulation of mutations could weaken this prediction.

\section{Acknowledgments}

I would like to thank: Maria Servedio, Joel Kingsolver, Troy Day,
Karin Pfennig, and Haven Wiley for scrutinizing my results; Sumit
Dhole, Alicia Frame, Caitlin Stern, Justin Yeh, Artur Romanchuk and
Daniel Promislow for providing helpful discussion and comments;
Richard M. Stallman and the GNU Project for guidance with software;
the developers of Bazaar, Trac, Guile, GNUPLOT and the GNU C Compiler
for creating reliable freedom-respecting software.  Samuel Tazzyman,
an anonymous reviewer and the editors of PeerJ provided helpful
comments that substantially improved the conceptual clarity of this
manuscript.

\bibliography{ss,multiple}

\end{document}